\documentclass[sigconf,screen,nonacm]{acmart}

\AtBeginDocument{%
  \providecommand\BibTeX{{%
    \normalfont B\kern-0.5em{\scshape i\kern-0.25em b}\kern-0.8em\TeX}}}

\setcopyright{rightsretained}

\acmConference[arXiv Preprint]{arXiv Preprint}{January 2026}{Adelaide, Australia}
\acmYear{2026}
\acmDOI{} 
\acmISBN{}

\acmSubmissionID{2075}

\usepackage{xcolor}
\usepackage{todonotes}
\usepackage{graphicx}
\usepackage{csquotes}
\usepackage{soul}

\begin{document}

\title{EndoExtract: Co-Designing Structured Text Extraction from Endometriosis Ultrasound Reports}



\author{Haiyi Li}
\email{a1949007@adelaide.edu.au}
\affiliation{%
  \institution{Univ. of Adelaide}
  \city{Adelaide}
  \state{SA}
  \country{Australia}
}

\author{Yiyang Zhao}
\email{yiyang.zhao@adelaide.edu.au}
\affiliation{%
  \institution{Univ. of Adelaide}
  \city{Adelaide}
  \state{SA}
  \country{Australia}
}

\author{Yutong Li}
\email{a1948101@adelaide.edu.au}
\affiliation{%
  \institution{Univ. of Adelaide}
  \city{Adelaide}
  \state{SA}
  \country{Australia}
}


\author{Alison Deslandes}
\email{alison.deslandes@adelaide.edu.au}
\affiliation{%
  \institution{Robinson Inst., Univ. of Adelaide}
  \city{Adelaide}
  \country{Australia}
}

\author{Jodie Avery}
\email{jodie.avery@adelaide.edu.au}
\affiliation{%
  \institution{Robinson Inst., Univ. of Adelaide}
  \city{Adelaide}
  \country{Australia}
}
\author{Mathew Leonardi}
\email{leonam@mcmaster.ca}
\affiliation{%
  \institution{McMaster University}
  \city{Hamilton}
  \country{Canada}
}
\author{Mary Louise Hull}
\email{louise.hull@adelaide.edu.au}
\affiliation{%
  \institution{Robinson Inst., Univ. of Adelaide}
  \city{Adelaide}
  \country{Australia}
}

\author{Hsiang-Ting Chen}
\email{tim.chen@adelaide.edu.au}
\affiliation{%
  \institution{Univ. of Adelaide}
  \city{Adelaide}
  \state{SA}
  \country{Australia}
}

\renewcommand{\shortauthors}{Li et al.}

\begin{abstract}
Endometriosis ultrasound reports are often unstructured free-text documents that require manual abstraction for downstream tasks such as analytics, machine learning model training, and clinical auditing.
We present \textbf{EndoExtract}, an on-premise LLM-powered system that extracts structured data from these reports and surfaces interpretive fields for human review. Through contextual inquiry with research assistants, we identified key workflow pain points: asymmetric trust between numerical and interpretive fields, repetitive manual highlighting, fatigue from sustained comparison, and terminology inconsistency across radiologists. These findings informed an interface that surfaces only interpretive fields for mandatory review, automatically highlights source evidence within PDFs, and separates batch extraction from human-paced verification. 
A formative workshop revealed that \textbf{EndoExtract} supports a shift from field-by-field data entry to supervisory validation, though participants noted risks of over-skimming and challenges in managing missing data.
\end{abstract}

\ccsdesc[500]{Human-centered computing~Graphical user interfaces}
\ccsdesc[300]{Computing methodologies~Natural language processing}
\ccsdesc[500]{Applied computing~Life and medical sciences}

\keywords{human-in-the-loop; large language models; clinical information extraction; interface design; medical reporting}

\maketitle

\section{Introduction}

Endometriosis assessment relies on specialist transvaginal ultrasound examinations that produce narrative reports describing lesions, anatomical sites, and measurements \citep{avery2024noninvasive}. Transforming these free-text reports into structured data for research and quality assurance remains labour-intensive and error-prone, requiring research assistants to manually transcribe findings into spreadsheets while navigating heterogeneous reporting styles and evolving templates.
\begin{figure*}[ht]
  \centering
  \includegraphics[width=0.9\textwidth]{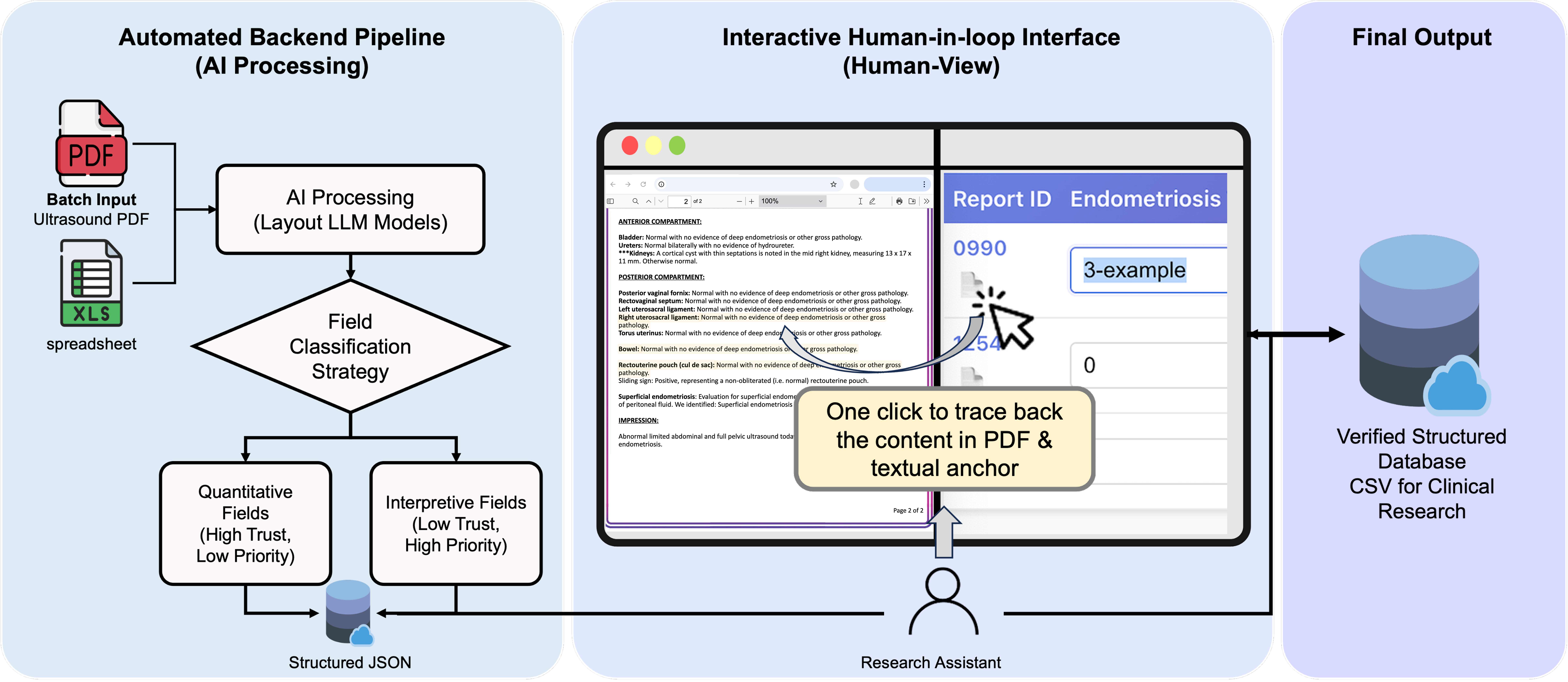}
  \caption{The EndoExtract Workflow. (Left) The backend utilizes a trust-based classification strategy to separate quantitative fields from interpretive ones. (Center) The clinician interface prioritizes human verification for low-trust/high-priority fields, featuring a one-click mechanism to anchor evidence in the original PDF. (Right) Verified data is exported for research.}
  \label{fig:system_workflow}
\end{figure*}
Earlier clinical information extraction approaches often assume standardised documentation templates or require substantial annotation effort \citep{Wang2018, castro2017automated, savova2017deepphe}. Large language models offer promising extraction capabilities for free-text reports, but healthcare deployment faces constraints from privacy requirements, data governance, and semantic failures that demand expert judgment \citep{beede2020human, agrawal2022large}. 
In addition, designing interfaces for LLM-based extraction presents additional challenges: effective human-AI collaboration requires understanding specific workflow contexts \citep{mackeprang2019discovering, rahman2022characterizing}, yet technical extraction research focuses on accuracy metrics rather than the situated realities of clinical data abstraction—where users navigate heterogeneous reporting styles, calibrate trust across field types, and manage attention under time pressure.

We address this gap with \textbf{EndoExtract}, an on-premise LLM-powered system co-designed with both specialist sonographers and research assistants who perform endometriosis report abstraction. The system treats the LLM as an accelerator for routine structuring rather than a fully autonomous extractor. The pipeline processes anonymised ultrasound report PDFs through schema-guided extraction with locally deployed LLMs, producing spreadsheet-ready outputs while preserving data privacy. The interface surfaces only interpretive fields requiring clinical judgment for mandatory review, while automatically highlighting source evidence within the original PDF to support verification. Batch processing separates extraction from review, allowing users to verify at their own pace. To our knowledge, \textbf{EndoExtract} is the first co-designed extraction interface for endometriosis ultrasound reporting that allocates human attention based on field-type trust asymmetry, embedding verification as the primary interaction model rather than an afterthought.

This paper contributes: (1) a contextual inquiry characterising workflow pain points, trust asymmetries, and verification practices among research assistants extracting data from endometriosis ultrasound reports; (2) \textbf{EndoExtract}, an on-premise LLM extraction system featuring selective review surfaces, automatic evidence highlighting, batch processing, and dual-version outputs for human review; and (3) design insights from a formative workshop validating the shift from manual data entry to supervisory validation and identifying evidence traceability, missingness management, and over-skimming constraints as key considerations for verification-centred human-AI collaboration.

\section{Contextual Inquiry}
We conducted contextual inquiry and think-aloud observation with 2 research assistants performing routine ultrasound scan report extraction at [anonymous Institution]. Sessions combined semi-structured interviews with direct workflow observation to identify recurring pain points, quality assurance practices, and trust patterns. All participants were experienced with gynecological ultrasound terminology and high-volume transcription demands. Sessions were audio-recorded with screen capture, with participant consent.

\textbf{Observed Work Models.}
Research assistants operate in a split-screen configuration with PDF reports alongside spreadsheets. Accessing source documents requires navigating 4–5 steps through VPN connections and shared drive hierarchies. Before extraction, assistants manually highlight critical diagnostic passages—lesion type, uterosacral nodules, obliterative changes, and bowel nodules—as visual anchors for cross-referencing. They then transcribe values while continuously comparing source and destination, maintaining verification logs for traceability. Reports range from brief sentences to 2,000-word narratives, demanding sustained concentration across variable document complexity.

\textbf{Key Findings and Breakpoints.}
Participants demonstrated asymmetric trust across field types: high confidence in objective numerical fields such as uterine dimensions and lesion volumes, but doubt toward interpretive or speculative language requiring clinical judgment. This trust asymmetry meant interpretive fields consistently required manual review despite automation potential. Routine highlighting of the same diagnostic passages before every extraction represented redundant effort unsupported by current tools. Prolonged close reading and comparison induced fatigue, with participants reporting degraded concentration and increased error risk over extended sessions. Fragmented navigation—multiple steps through VPN and folder structures—repeatedly interrupted the side-by-side workflow critical for verification. Finally, terminology inconsistency across radiologists (e.g., "Douglas pouch" variations; "lesion" versus "mass" versus "nodule") added cognitive load and introduced potential for extraction inconsistency.

\textbf{Design Implications}
These findings shaped three core design decisions. First, because participants trusted quantitative values but doubted interpretive text, the interface directs human attention toward interpretive fields while automating trusted numerical extraction. Second, embedding source evidence highlighting directly within the interface eliminates the redundant manual marking observed in current practice. Third, separating automated extraction from human review enables paced, interruptible workflows that mitigate fatigue from sustained high-intensity comparison.

\begin{figure*}[h]
  \centering
  \includegraphics[width=0.9\linewidth]{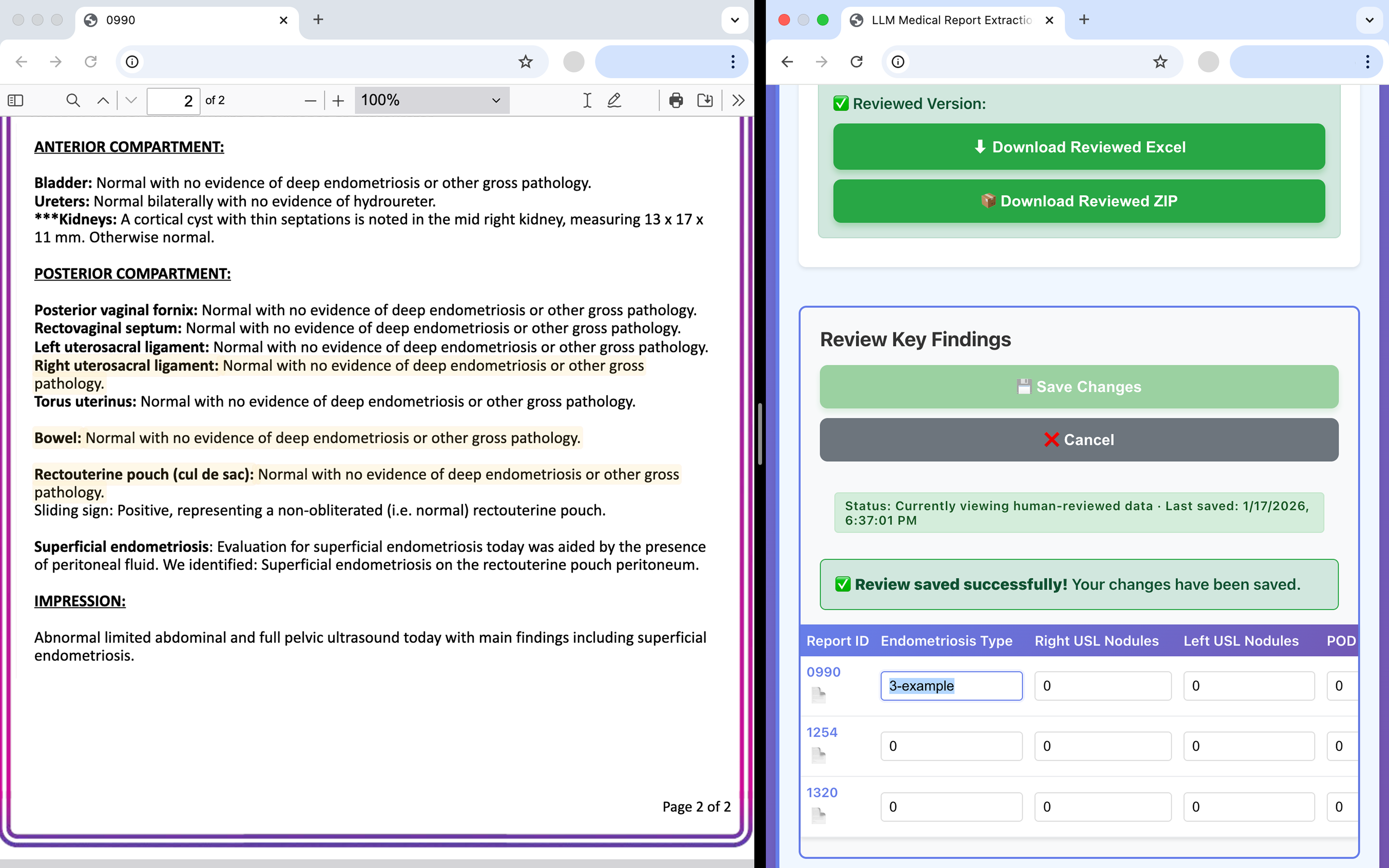}
  \caption{Selective review interface illustrating how highlighted evidence excerpts are shown alongside extracted fields to support human review.}
    \label{fig:system_feature}
\end{figure*}
\section{EndoExtract}

\textbf{EndoExtract} is an extraction system that supports human review, directly informed by the contextual inquiry findings. The workflow consists of batch extraction followed by human-paced review of select fields. Users upload PDF reports for background processing; upon completion, key fields appear in a unified table awaiting mandatory human review. Complete extraction outputs remain available as downloadable structured files. Figure~\ref{fig:system_workflow} shows the complete workflow. The extraction backend is powered by \textbf{gpt-oss-20b}, a locally deployed model running via Ollama. We selected this architecture to balance the reasoning capability required for interpreting ambiguous medical phrasing with the strict constraint of on-premise processing, ensuring that sensitive patient data never leaves the secure institutional environment. With the backend privacy secured, the frontend focuses on the human verification experience. Based on the contextual inquiry, \textbf{EndoExtract} incorporates several unique features in the UI design (Figure~\ref{fig:system_feature}):

\textbf{Selective Review Surface.} The contextual inquiry revealed asymmetric trust: research assistants confidently accepted numerical values but consistently questioned interpretive content requiring clinical judgment. Accordingly, \textbf{EndoExtract} exposes only five interpretive fields for mandatory human review—Endometriosis Type, Right Uterosacral Ligament Nodules, Left Uterosacral Ligament Nodules, Pouch of Douglas Obliteration, and Bowel Deep Infiltrating Endometriosis. These fields remain inline-editable with immediate visual feedback. The remaining 150+ quantitative fields are extracted automatically, with optional access for editing when needed. This selective surfacing directs human attention where clinical judgment is essential while preserving efficiency for trusted numerical data.

\textbf{Evidence Highlighting.}
Research assistants routinely highlighted the same diagnostic passages before every extraction—a redundant manual step unsupported by existing tools. \textbf{EndoExtract} eliminates this repetition by storing evidence sentences during extraction and displaying them automatically. A one-click "Open PDF" action presents highlighted source passages alongside the editable review table, preserving the split-screen spatial workflow participants preferred while removing the 4–5 navigation steps previously required to access source documents through VPN and folder hierarchies.

\textbf{Batch Processing and Paced Review.}
Participants reported fatigue from sustained high-intensity comparison across variable-length reports, with degraded concentration increasing error risk. \textbf{EndoExtract} addresses this by separating automated extraction from human review. The upload interface accepts up to 5,000 PDF files for background processing with progress indicators, freeing users to continue other tasks. Upon completion, all key results appear in a unified table view where users review at self-determined pace. Change tracking indicates modification status, and batch save commits all changes simultaneously to both the review table and the human-modified Excel output.


\textbf{Semantic Normalisation.}
The contextual inquiry identified terminology inconsistency as a source of cognitive load and extraction error—different radiologists label identical structures variably (e.g., "POD obliteration" versus "obliterated pouch of Douglas"). \textbf{EndoExtract} addresses this through automated semantic normalisation, mapping synonymous anatomical terms and mixed units to consistent labels during extraction. This reduces interpretive burden and improves consistency in downstream analysis

\section{Formative Workshop Evaluation}
We conducted a formative workshop to evaluate whether the interface supports human review in practice. The workshop comprised two activities: (1) A \textbf{think-aloud walkthrough} in which the participant completed the extraction-to-review loop of one ultrasound scan report while verbalising intentions, cues noticed, and reactions to interface elements. We logged in-situ responses to evidence highlighting, split-view comparison, and field confirmation or editing decisions. This activity evaluated whether critical review entry points were salient and whether evidence-finding and correction could be performed with low friction. (2) A \textbf{collaborative reflection interview} examined workflow transformation, trust and verification strategies, ambiguity handling, and adoption concerns.

\textbf{Participants.} We conducted the workshop with four domain experts from the [anonymous Institution] to evaluate the system's clinical validity. The panel included a specialist sonographer (doctoral level), a senior epidemiologist with extensive experience in women’s health research, a health technology assessment analyst, and a clinical research coordinator involved in endometriosis studies. This composition ensures that our evaluation captures perspectives from clinical domain knowledge, operational workflow, and data quality assurance.

\subsection{Workshop Findings}

The workshop leads to the following finding:

\textbf{Workflow transformation.} The participant framed the experience as "checking and correcting" rather than entering data from scratch, suggesting a role shift from field-by-field data entry to supervisory validation of AI outputs. Notably, verifying numeric fields remained effortful due to semantic mapping back to report context, whereas reading text with highlighted evidence made locating information and making judgments easier. The system reduces workload primarily by simplifying evidence retrieval for interpretive fields rather than equally reducing effort across all field types.

\textbf{Trust and verification strategies.} The participant indicated that full item-by-item checking was unnecessary: with highlighted evidence, they could skim certain sections while focusing attention on fields requiring judgment. Evidence highlighting enabled rapid location of supporting text, increasing confidence in confirmation and editing decisions. This skim-enabled review depends on evidence reliability and traceability—highlighting functions as a trust calibration mechanism.

\textbf{Ambiguity and atypical cases.} Ambiguous phrasing (e.g., "possible," "suspicious for") varies across clinicians and contexts, requiring human interpretation at this stage. The participant suggested that such terms could potentially be standardised by teaching the model consistent mappings, noting that an LLM might achieve greater consistency than human reviewers who apply different thresholds to identical phrases. However, atypical clinical presentations—where conditions lack characteristic patterns—require experience-driven judgment that resists binarisation. The system must preserve explainable and reversible review pathways for uncertain cases.

\textbf{Adoption concerns.} The participant preferred leaving fields blank over filling them with unsupported values, viewing blankness as an honest signal when evidence is absent. They requested that empty fields be proactively marked to avoid additional effort locating missing items. Hallucination and false reassurance emerged as primary adoption risks—interface cues and automated checks could create false security encouraging excessive skimming and missed errors. Technical friction in LLM deployment was also flagged as an organisational barrier.
\section{Discussion}
Our findings surface three design principles for verification-centred human-AI collaboration in clinical data abstraction:

\textbf{Guided Attention Allocation.} Our inquiry revealed that user trust depends on the data type: users trusted extracted numbers but doubted interpretive descriptions. 
We avoided displaying fine-grained machine learning model confidence scores, which would force users to constantly evaluate model uncertainty. 
Instead, the interface uses field type to direct attention, requiring manual review only for interpretive fields while automating low-risk numerical data to save effort.

\textbf{Asynchronous Human-AI Collaboration.} 
While real-time iterative human-AI interaction offers agency, it often imposes unnecessary cognitive burden through frequent interruptions. We addressed this by separating extraction from verification. 
This design allows the system to operate at machine speed while users review batched results at their own pace. 
Ultimately, this shifts efficiency from simple task acceleration to user-controlled workflow scheduling, transforming the AI from a potentially disruptive tool into a flexible collaborator that respects human availability.

\textbf{Evidence as a Reasoning Trace.}
Manual verification often devolves into a tedious search for information within a document. 
We transformed this into a seamless validation task by visualizing the system's "reasoning trace" via automatic highlighting. 
Crucially, we displayed this evidence in the original PDF rather than embedding it in the table. This approach maintains the full narrative and spatial context of the clinical report, allowing users to cross-reference findings without confusing AI predictions with the original medical records.

\section{Limitations and Future Work}
Several limitations warrant consideration. The categorisation of fields into interpretive and quantitative types was derived from a specific endometriosis ultrasound workflow and may not generalise to other clinical domains with different trust asymmetries. The design assumes stable trust patterns, but verification behaviour will likely evolve with experience and vary across individuals. While evidence highlighting enables selective review, it introduces over-skimming risk—workshop findings identified false reassurance as a primary adoption concern, where interface cues could encourage users to skim critical fields and miss errors. The system does not yet implement mechanisms to constrain over-skimming or explicitly manage ambiguous phrasing and atypical cases that resist binarisation. 

\section{Conclusion}
\textbf{EndoExtract} demonstrates a verification-centred approach to LLM-assisted clinical data extraction. Grounded in contextual inquiry, the system allocates human attention to interpretive fields through selective review surfaces, eliminates redundant highlighting through automatic evidence linking, and mitigates fatigue through batch processing with paced review. Workshop evaluation confirmed a shift from manual data entry to supervisory validation, with evidence highlighting enabling efficient selective review. Three generalisable insights emerged: evidence-highlighting as implicit explanation, asynchronous human-AI collaboration that yields efficiency through user-controlled pacing, and guided attention allocation as an alternative to uniform automation or exhaustive verification. Future work will evaluate workflow impact in situ and explore generalisation across structured abstraction domains.


\bibliographystyle{ACM-Reference-Format}
\bibliography{acmart}

\appendix

\end{document}